\documentclass[aps,prl,twocolumn,superscriptaddress]{revtex4-2}
\usepackage{graphicx}
\usepackage{dcolumn}
\usepackage{bm}
\usepackage{physics}

\begin{document}
\setcounter{secnumdepth}{2}

\title{Bounds on the Entropy Production in Stochastic Resetting with Harmonic Traps}
\author{Oded Farago}
\affiliation{Department of Biomedical Engineering,
Ben-Gurion University of the Negev,
Be'er Sheva 85105, Israel}
\affiliation{Institute of Physics and Astronomy,
University of Potsdam,
Potsdam-Golm 14476, Germany}
\affiliation{Max Planck Institute of Colloids and Interfaces, Potsdam-Golm 14476, Germany}
\date{\today}

\begin{abstract}
We study the entropy production associated with stochastic resetting of a Brownian particle in an optical trap, where resetting is implemented by increasing the trap stiffness to restore a localized distribution. We interpret the dynamics as a thermodynamic cycle consisting of thermal expansion during diffusion and compression performed by an external resetting agent. Within this framework, we distinguish between the irreversibility of the resetting cycle and the additional entropy production associated with its physical implementation. We establish a lower bound based on the Kullback–Leibler (KL) divergence between forward and reversed trajectory ensembles. For a harmonic trap, this divergence can be expressed entirely in terms of the probability distributions before and after resetting, independently of the resetting protocol. An upper bound is obtained by adding the operational entropy production associated with instantaneous compression, quantified by its excess work relative to quasistatic resetting, to the KL contribution. For free diffusion between resetting events, we derive explicit analytical expressions for both contributions to the entropy-production rate as functions of the resetting rate. Remarkably, they approach the same finite value in the limit of vanishing resetting rate, as the increasing entropy production per cycle compensates for the decreasing frequency of resetting events. At high resetting rates, both contributions vanish as the particle remains effectively confined to its initial distribution, with little time to spread between resets.
\end{abstract}

\maketitle

\section{Introduction}

Stochastic resetting refers to a class of nonequilibrium processes in which a system evolving under stochastic dynamics is intermittently returned to a prescribed state, typically its initial condition, at random times. This modification, introduced in the context of diffusive search by Evans and Majumdar~\cite{EvansMajumdar2011PRL}, has profound consequences for both the transient and steady-state behavior of stochastic systems~\cite{Evans2020JPA}. In contrast to standard diffusion, where probability distributions broaden indefinitely, resetting induces a nontrivial stationary state even in unbounded domains, characterized by non-Gaussian distributions and persistent probability currents~\cite{EvansMajumdar2011JPA, EuleMetzger2016NJP, Pal2016JPA}. A central motivation for introducing resetting is its effect on first-passage properties. Resetting renders otherwise divergent mean first-passage times finite by curbing long unproductive excursions away from the target~\cite{EvansMajumdar2011PRL, EvansMajumdar2011JPA, Kusmierz2014PRL}, and optimal resetting rates exist across a broad class of processes~\cite{KusmierzGudowskaNowak2015PRE, Reuveni2016PRL, PalReuveni2017PRL, PalKostinskiReuveni2022JPA}. The maintenance of the resulting nonequilibrium steady states necessarily involves entropy production, and quantifying this irreversibility is the central concern of the present work.

Beyond its applications to first-passage problems, stochastic resetting has stimulated a range of theoretical developments. The characteristic exponential or Laplace profiles of its stationary distributions have been investigated~\cite{EvansMajumdar2011JPA, Majumdar2015PRE, Pal2016JPA}, and the framework has been extended to diffusion in potentials and higher dimensions~\cite{EvansMajumdar2014JPA, PAL2015PRE}. The renewal structure induced by resetting allows exact formulations and has been exploited to analyze time-dependent properties and relaxation to steady states~\cite{EvansMajumdar2011JPA, EuleMetzger2016NJP, PalReuveni2017PRL}. Resetting has also been used to study large-deviation properties and dynamical phase transitions~\cite{Meylahn2015PRE, HarrisTouchette2017JPA}, as well as extensions involving non-Poissonian resetting, space-dependent protocols, and resetting with memory~\cite{BoyerSolis2014PRL, EuleMetzger2016NJP, ChechkinSokolov2018PRL}.

Despite this extensive body of work, the thermodynamic characterization of stochastic resetting remains comparatively less developed, and several conceptual difficulties have limited progress. A central challenge is that the standard formulation of stochastic thermodynamics relies on comparing forward trajectories with their time-reversed counterparts. In resetting processes, reset events are intrinsically unidirectional and do not admit natural reverse transitions, which obstructs a direct application of this framework~\cite{Rahav2017PRE,Busiello2020PRR}. Early work by Fuchs, Goldt, and Seifert~\cite{Fuchs2016EPL} addressed this difficulty by adopting an augmented Fokker-Planck description that explicitly accounts for the nonlocal probability flux associated with reset events. They identified a Landauer-type entropy contribution~\cite{Landauer1961IBM} associated with information erasure at the reset location. However, perfect resetting to a delta function introduces a mathematical singularity that renders the trajectory-level entropy production ill-defined~\cite{Busiello2020PRR}. Mori, Olsen, and Krishnamurthy~\cite{Mori2023PRR} overcame this difficulty by replacing the delta-function reset with a spread distribution $p_R(x)$ nonzero everywhere, thereby restoring a well-defined time-reversed process. The resulting entropy production takes the form of a symmetric Kullback-Leibler (KL) divergence between the steady state and the resetting distribution. Despite this formal advance, their description still does not account for the physical cost of compressing the distribution during resetting.

Experimentally, stochastic resetting is most naturally realized in optical trapping setups, where a colloidal particle is confined by a laser-controlled harmonic potential and resetting is implemented by physically compressing the distribution through a rapid increase of the trap stiffness~\cite{TalFriedman2020JPCL,Besga2020PRR,Goerlich2025PRE,Sherf2026PRE}. As explained above, this physical picture is not captured by the teleportation model underlying the approaches of refs.~\cite{Fuchs2016EPL,Mori2023PRR}. To identify the operational contribution associated with the resetting protocol, we first consider how entropy production is quantified within the general framework of stochastic thermodynamics. We begin by recalling that entropy is a well-defined thermodynamic state function only at equilibrium, and that the Shannon entropy coincides with this state function only under these conditions~\cite{Parrondo2009NJP}. Away from equilibrium, the Shannon entropy characterizes the spread of the distribution but is not directly related to heat or entropy exchange with the environment. In stochastic thermodynamics, total entropy production is quantified by the KL divergence between the probability of a forward trajectory and its time-reversed counterpart, a framework rooted in fluctuation theorems~\cite{Jarzynski1997PRL,Crooks1999PRE,Seifert2005PRL,Kawai2007PRL,GomezMartin2008PRE}. To connect this formulation with the thermodynamic definition of entropy, the process must be coupled to thermal reservoirs at its endpoints, so that the initial and final states are equilibrium states for which thermodynamic entropy is well defined~\cite{Parrondo2009NJP,Esposito2010PRE}. The trajectory-level KL divergence then decomposes into a state contribution determined by the initial and final distributions, and an irreversible contribution generated during the transformation~\cite{Roldan2012PRE}.

In optical trap realizations of stochastic resetting, the temperature remains constant throughout the cycle. Equilibration at the beginning and end of each cycle can, in principle, be achieved by adjusting the trap stiffness, allowing the particle to reach equilibrium states corresponding to different stiffness values~\cite{Rosales-Cabara2020PRR,Olsen2024PRR}. Goerlich et al.~\cite{Goerlich2025PRE} characterized the thermodynamic cost of such a cycle by casting the optical trap as an external Maxwell demon, whose act of compressing and releasing the particle distribution converts information into work required to operate the resetting cycle. The demon’s work is identified with the work associated with varying the trap stiffness~\cite{Sekimoto1998}, which is increased to large values to compress the distribution to the reset state. Building on the nonequilibrium Landauer principle~\cite{Esposito2011EPL}, they implement a Carnot-like protocol that achieves a lower bound on the demon’s work per cycle. In this protocol, at each reset event, the stiffness is first jumped instantaneously to bring the system to equilibrium, and then increased quasistatically and reversibly to the reset value. The excess irreversible contribution to the entropy production therefore vanishes in this protocol, reducing the entropy production to the endpoint KL expression. As with the Carnot bound for heat engines, however, achieving this reversible limit requires quasistatic operation, making it a thermodynamic benchmark rather than a practical protocol for resetting at finite rates.

Here, we consider a reset protocol in which the stiffness is switched instantaneously to the large value corresponding to the localized reset state. Excluding protocols where the stiffness overshoots above this value and subsequently reduced, this is the opposite extreme to the quasistatic protocol presented in~\cite{Goerlich2025PRE}. While the quasistatic protocol maintains the distribution in equilibrium throughout the stiffness increase, the instantaneous switch couples the system immediately to the final localized potential, so that the distribution is maximally displaced from its new equilibrium at the onset of the compression and relaxes toward it throughout. Since the compression speed is set by the trap stiffness and the particle mobility, and both can in principle be made arbitrarily large, this protocol provides the closest physical realization of the teleportation limit. We investigate the irreversible entropy production associated with stochastic resetting in an optical trap. We first show that, for a harmonic trap, the trajectory-level KL divergence over a complete resetting cycle reduces exactly to an expression involving only the endpoint distributions, independently of the compression protocol. This provides a protocol-independent lower bound on the entropy production. We then consider the additional entropy production associated with the physical operation of the trap, which depends on the stiffness-variation protocol and allows us to establish an upper bound on the total entropy production.

\section{The Thermodynamics of Resetting Cycles}
\subsection{Setup and Dynamics}

We consider the overdamped dynamics of a Brownian particle moving in a one-dimensional potential $V(x)$ with a damping coefficient $\gamma$. In the intervals between resetting events, the time evolution of the particle position $x(t)$ is described by the overdamped Langevin equation
\begin{equation}
\gamma \dot{x} = -\frac{\partial V(x)}{\partial x} + \sqrt{2\gamma k_B T} \xi(t),
\label{eq:lang}
\end{equation}
where $\xi(t)$ is Gaussian white noise with zero mean and a correlation function $\langle \xi(t) \xi(t') \rangle = \delta(t-t')$. Equivalently, the evolution of the probability density $P(x,t)$ is governed by the continuity equation
\begin{equation}
\frac{\partial P(x, t)}{\partial t} = -\frac{\partial J(x, t)}{\partial x},
\label{eq:coneq}
\end{equation}
where the probability current $J(x,t)$ is defined as
\begin{equation}
J(x, t) = -\frac{1}{\gamma} \left[ \frac{\partial V(x)}{\partial x} P(x, t) + k_B T \frac{\partial P(x, t)}{\partial x} \right] .
\label{eq:fokplan}
\end{equation}

The resetting protocol  consists of a spreading stage, followed by compression that restores the initial localized distribution through harmonic confinement. We focus on free diffusion during the expansion stage, while the case of diffusion in a harmonic potential of finite stiffness $V(x)=\frac{1}{2}kx^2$ is considered in Appendix~\ref{app:finite_stiffness}. In both cases, an initially Gaussian distribution, $P(x,0)=\mathcal{N}(0,\sigma_0^2)$, remains Gaussian at all times:
\begin{equation}
P(x,t)=\frac{1}{\sqrt{2\pi\sigma^2(t)}}\exp\left(-\frac{x^2}{2\sigma^2(t)}\right),
\label{eq:Gauss}
\end{equation}
with variance $\sigma^2(t)=\sigma_0^2+2Dt$ for free diffusion, where $D=k_BT/\gamma$, and $\sigma^2(t)=k_BT/k+(\sigma_0^2-k_BT/k)\exp(-2kt/\gamma)$ for diffusion in a harmonic potential of stiffness $k$. At the onset of resetting, the Gaussian distribution with variance $\sigma^2(t)$ corresponds to a thermal equilibrium state of a harmonic oscillator with stiffness $k(t)=k_BT/\sigma^2(t)$. The trap stiffness is then instantaneously increased to $k_0=k_BT/\sigma_0^2$, corresponding to the target localized state of variance $\sigma_0^2$. Since harmonic confinement preserves Gaussianity, the distribution remains Gaussian during compression. In an ensemble sense, resetting therefore reverses the spreading of the distribution, driving its variance from $\sigma^2(t)$ back to $\sigma_0^2$ along the same family of Gaussian distributions. These distributions can be identified with equilibrium states of harmonic oscillators with stiffnesses ranging from $k(t)$ to $k_0$. The speed of this process depends on the particle mobility and the applied trap stiffness. In optical trapping experiments, the minimal achievable localization width $\sigma_0^2$ is set by the stiffness of the trap and the sampling rate of the control protocol~\cite{TalFriedman2020JPCL,Besga2020PRR}. At a more fundamental level, the applicability of the classical description requires the localization scale to remain sufficiently large compared with the thermal de Broglie wavelength $\lambda_{dB}=h/\sqrt{2\pi m k_BT}$.

\subsection{Entropy Production: General Framework}
\label{subsec:general}

As described above, the resetting cycle connects two equilibrium states with stiffnesses $k(t)$ and $k_0$. The entropy production for such a process can generally be decomposed into two contributions~\cite{Parrondo2009NJP}. The first is the equilibrium entropy difference associated with the initial and final ensembles, while the second is the irreversible entropy generated during the process. Applying this decomposition to the compression stage of the resetting protocol, we denote the two contributions by $\Delta S_{\rm state}$ and $\Delta S_{\rm irr}$, respectively. The total entropy production, including both contributions, can also be
expressed as a KL divergence between the path probability densities
$P[x]$ and $\tilde P[x]$ associated with the forward and time-reversed
trajectories,
\begin{eqnarray}
\label{eq:KLD}
\Delta S
&=&
\Delta S_{\rm state}
+
\Delta S_{\rm irr} \\
&=&
k_B D(P\|\tilde P)
=
k_B \int \mathcal{D}x\, P[x]\ln\left(\frac{P[x]}{\tilde P[x]}\right),\nonumber
\end{eqnarray}
where $\mathcal{D}x$ denotes integration over trajectories. This quantity
is non-negative, which is a manifestation of the second
law~\cite{Parrondo2009NJP,Kawai2007PRL}.

The exact expression in Eq.~(\ref{eq:KLD}) is formulated in terms of trajectory probability densities and therefore requires complete information about the stochastic evolution. In practice, however, thermodynamic descriptions are often constructed only from the endpoint probability distributions, without explicit knowledge of the intermediate trajectories. This reduction from path ensembles to endpoint distributions discards information about the intermediate states, meaning the inferred entropy production generally underestimates the exact trajectory-based value~\cite{GomezMartin2008PRE,Roldan2012PRE}. The corresponding KL divergence constructed from the initial and final probability distributions \(P(x,t_i)\) and \(P(x,t_f)\),
\begin{equation}
\Delta S_{\rm KL}
=
k_B\int dx\, P(x,t_i)
\ln\left(\frac{P(x,t_i)}{P(x,t_f)}\right),
\label{eq:deltaskl}
\end{equation}
is a state-dependent quantity associated with the coarse-grained nonequilibrium process. This suggests an alternative decomposition of the total entropy
production into the endpoint KL contribution and a non-negative
excess term,
\begin{equation}
\Delta S
=
\Delta S_{\rm KL}
+
\Delta S_{\rm excess},
\qquad
\Delta S_{\rm excess}\ge0.
\label{eq:KLandextra}
\end{equation}
The excess term encompasses two distinct sources of entropy
production: (i) the dynamical contribution, which reflects
information about the intermediate trajectories that is lost
when only the endpoint distributions are retained, and
(ii) the operational contribution, which arises from dissipation
associated with the implementation of the compression protocol.
Both, in principle, may  depend on the specific protocol used to restore
the localized state.

To characterize the entropy production over a thermodynamic cycle,
we must also consider the reverse transformation from the localized
state to the thermally expanded state reached at the end of the
spreading stage, immediately before the compression begins. Summing the entropy productions associated with these two transformations, the equilibrium state contributions in Eq.~(\ref{eq:KLD}) cancel over the cycle, leading to a symmetric KL divergence,
\begin{equation}
\Delta S_{\rm cycle}
=
k_B\int \mathcal{D}x\,
\left(P[x]-\tilde P[x]\right)
\ln\left(\frac{P[x]}{\tilde P[x]}\right).
\label{eq:symmetricKL}
\end{equation}
This quantity provides a non-negative measure of the irreversible entropy generated over the thermodynamic cycle. The corresponding endpoint expression, obtained by replacing the path probability densities with the initial and final distributions, provides a lower bound
\begin{equation}
\Delta S_{\rm KL,cycle}
\!=\!
k_B\!\int\!dx\, \left[P(x,t_i)-P(x,t_f)\right]
\ln\left(\frac{P(x,t_i)}{P(x,t_f)}\right).
\label{eq:deltasklcycle}
\end{equation}
For the Gaussian distributions of Eq.~(\ref{eq:Gauss}),
the cycle KL divergence in Eq.~(\ref{eq:deltasklcycle})
can be evaluated explicitly, yielding
\begin{equation}
\Delta S_{\rm cycle} \geq \Delta S_{\rm KL,cycle}
=
\frac{k_B}{2}
\left[
\frac{\sigma^2(t)}{\sigma_0^2}
+
\frac{\sigma_0^2}{\sigma^2(t)}
-2
\right].
\label{eq:deltascycle}
\end{equation}

\subsection{Dynamical contribution}
\label{subsubsec:dtnamical}

We now examine the dynamical contribution to the entropy
production in the harmonic resetting cycle and show that
it vanishes independently of the compression protocol.
Consider a trajectory starting at $(x_i,t_i)$ and ending
at $(x_f,t_f)$. For a Markov process, the path probability
factorizes into a product of infinitesimal propagators,
so that the logarithm of the forward-to-reverse trajectory
probability ratio is
\begin{equation}
\ln\frac{P[x]}{\tilde P[x]} = \ln\frac{P(x_i,t_i)}{P(x_f,t_f)} + \sum_n \ln\left[\frac{G(x_{n+1},t_{n+1}|x_n,t_n)}{G(x_n,t_n|x_{n+1},t_{n+1})}\right].
\label{eq:propagator}
\end{equation}
This separation into endpoint and dynamical terms parallels
the decomposition in Eq.~(\ref{eq:KLandextra}), with the
first term corresponding to $\Delta S_{\rm KL}$. Since it is evaluated at the initial time, the ensemble average is taken with respect to $\mathcal{N}(0,\sigma_i^2)$, giving
\begin{eqnarray}
\left\langle\ln\frac{P(x_i,t_i)}{P(x_f,t_f)}\right\rangle &=& \left\langle -\frac{x_i^2}{2\sigma_i^2} + \frac{x_f^2}{2\sigma_f^2}\right\rangle_{\mathcal{N}(0,\sigma_i^2)} + \frac{1}{2}\ln\left(\frac{\sigma_f^2}{\sigma_i^2}\right) \nonumber\\
&=& -\frac{1}{2} + \frac{\sigma_i^2}{2\sigma_f^2} + \frac{1}{2}\ln\left(\frac{\sigma_f^2}{\sigma_i^2}\right).
\end{eqnarray}
Adding the contribution from the reverse branch, evaluated with respect to $\mathcal{N}(0,\sigma_f^2)$, the logarithmic terms cancel and the total endpoint contribution to the symmetric KL divergence is precisely $\Delta S_{\rm KL,cycle}$ of Eq.~(\ref{eq:deltascycle}), with $\sigma_i=\sigma_0$ and $\sigma_f=\sigma(t)$. 

We now turn to the dynamical contribution to
$\Delta S_{\rm excess}$ in Eq.~(\ref{eq:KLandextra}),
represented by the second term on the r.h.s.\
of Eq.~(\ref{eq:propagator}), and show that it
vanishes over the full resetting cycle.
The harmonic potential is special because the evolving
distribution remains Gaussian throughout, confined to
the manifold parametrized by $\sigma^2(t)$.
Using the Stratonovich convention, each propagator ratio
in the sum is evaluated at the midpoint variance
$\sigma^2_{n+1/2}$, yielding
\begin{equation}
\ln\left[
\frac{G_{\sigma^2_{n+1/2}}(x_{n+1}|x_n)}
{G_{\sigma^2_{n+1/2}}(x_n|x_{n+1})}
\right]
=
-\frac{x_{n+1}^2-x_n^2}{2\sigma^2_{n+1/2}}.
\end{equation}
Passing to the continuum limit and taking the ensemble
average, the propagator contribution for each
transformation becomes
\begin{equation}
\left\langle -\int\frac{x\,dx}{\sigma^2(t)}\right\rangle
=
-\int_{\sigma_i^2}^{\sigma_f^2}
\frac{d\sigma^2}{2\sigma^2}
=
\frac{1}{2}\ln\left(\frac{\sigma_i^2}{\sigma_f^2}\right).
\end{equation}
For the spreading and compression stages, the initial
and final variances exchange their roles. Consequently,
the two logarithmic terms have equal magnitudes and
opposite signs, and their sum vanishes over the full
resetting cycle. This cancellation is independent of
the time dependence of the variance and therefore of
the specific compression protocol. It reflects the
fact that the dynamical information relevant to entropy
production is fully captured by the endpoint variances
alone. This leaves us with only the operational
contribution to $\Delta S_{\rm excess}$, associated
with the implementation of the compression protocol.

\subsection{Operational contribution}

The operational contribution to entropy production can be
analyzed within the external demon framework introduced
in~\cite{Goerlich2025PRE}. The resetting agent acts as a demon that reduces the positional entropy of the particle. It does so by compressing the Gaussian distribution, which has spread to variance $\sigma^2(t)$ after a diffusion time $t$, back to the equilibrium Gaussian at stiffness $k_0$ with variance $\sigma_0^2 = k_BT/k_0$. To do so, the demon must invest external energy, with an average work given by~\cite{Sekimoto1998}
\begin{equation}
\langle w^{\rm ext}\rangle = \frac{1}{2}\int_{t_i}^{t_f} \dot{k}(s)\sigma^2(s)\,ds,
\label{eq:sakiwork}
\end{equation}
where $t_i=t$ and $t_f=t+t^*$ denote, respectively, the beginning and end of the compression stroke whose duration is $t^*$, and $k(s)$ ($t_i\leq s\leq t_f$) is the resetting protocol.

Let us first consider the quasistatic protocol~\cite{Goerlich2025PRE}, in which the stiffness is first increased instantaneously to the value $k(t)=k_BT/\sigma^2(t)$, for which the expanded distribution is an equilibrium distribution. The stiffness is then increased quasistatically from $k(t)$ to $k_0$, keeping the system in equilibrium throughout this compression stage. For this protocol, the operational contribution is already
accounted for by the trajectory-level KL divergence,
given by Eq.~(\ref{eq:KLD}) for each branch and by its
symmetric form, Eq.~(\ref{eq:symmetricKL}), for the full cycle.
This is because the KL expression also accounts for the
entropic cost of restoring equilibrium at the end of a
transformation. In the present protocol, this is achieved
by switching the trap stiffness to $k(t)$ at the end of
the diffusion stage. The subsequent quasistatic compression is reversible and therefore produces no additional entropy. Consequently, the quasistatic resetting protocol introduces no operational entropy production beyond that already accounted for by the trajectory-level KL divergence. As we already established above in section~\ref{subsubsec:dtnamical}, for the harmonic potential $\Delta S_{\rm cycle} = \Delta S_{\rm KL,cycle}$, implying that the total entropy production for the quasistatic cycle is given by Eq.~(\ref{eq:deltascycle}).  

For a general resetting protocol, the cycle can still be divided into two processes connecting the equilibrium distributions corresponding to the harmonic trap stiffnesses $k_0$ and $k(t)$. The first begins with the localized Gaussian distribution
of variance $\sigma_0^2$. The trap is then switched off
($k_0\rightarrow 0$), allowing the distribution to spread
freely until its variance reaches $\sigma^2(t)$, at which
point the equilibrium stiffness is introduced
[$0\rightarrow k(t)$]. The second consists of compression, during which the stiffness is increased from $k(t)$ to $k_0$ at a finite rate. Finite-rate compression generally requires the external demon to perform additional work, which is dissipated as frictional heat into the thermal bath during the compression stage. To quantify the associated contribution to the entropy production, we first need to calculate the extra work required from the external demon
\begin{equation}
\Delta W=\langle w^{\rm ext}\rangle-\langle w^{\rm ext}\rangle_{\rm qs},
\label{eq:extrawork}
\end{equation}
where $\langle w^{\rm ext}\rangle_{\rm qs}$ is the work performed in the quasistatic reference protocol. Using Eq.~(\ref{eq:sakiwork}), and recalling that for quasistatic compression $k(s)\sigma^2(s)=k_BT$, we find that  
\begin{equation}
\langle w^{\rm ext}\rangle_{\rm qs}=\frac{k_BT}{2}\ln\left(\frac{\sigma^2(t)}{\sigma_0^2}\right).
\end{equation}
Notice that this result does not include the initial work done when the trap is switched on, which establishes the equilibrium reference state and is accounted for as part of the diffusion branch. 
For the instantaneous switch directly from $k(t)$ to $k_0$ [$k(s)=k(t)+[k_0-k(t)]H(s-t)$, where $H$ is the Heaviside function]
\begin{equation}
    \Delta W^{\rm max}=\frac{k_BT}{2}\left[\frac{\sigma^2(t)}{\sigma_0^2}-1-\ln\left(\frac{\sigma^2(t)}{\sigma_0^2}\right)\right],
    \label{eq:exworkinst}
\end{equation}
where the superscript "max" stands for the fact that this is the maximal excess work over all the protocols in which the stiffness increases monotonically from \(k(t)\) to \(k_0\). 

The excess entropy production associated with $\Delta W$ raises a fundamental conceptual question concerning the choice of thermodynamic system boundaries. So far, we have considered a system comprising the particle, the heat bath at temperature $T$, and the external demon controlling the trap stiffness. The energy source from which the demon draws the energy required to perform its work has been excluded from this description. The excess work $\Delta W$ is eventually dissipated into the thermal bath, contributing an additional entropy $\Delta W/T$. However, including the energy source in the thermodynamic accounting may introduce either negative or positive entropy contributions. For example, if energy is drawn from an external heat reservoir at temperature $T^{\rm ext}>T$, the reservoir loses entropy, reducing the net contribution. Conversely, a dissipative energy source, such as a battery with internal resistance, may generate additional entropy while supplying energy to the demon. Such contributions depend on the physical implementation of the resetting mechanism and cannot be determined from the particle dynamics alone. 

To avoid introducing implementation-dependent considerations that lie outside the scope of our theoretical description, we treat the demon as being powered by an ideal work source, which supplies energy without an accompanying entropy change. Within this framework, the only operational entropy production arises from the excess work which is dissipated to the bath $\Delta S_{\rm op}=\Delta W/T$. This contribution vanishes for the quasistatic protocol, for which the total entropy production is given by $\Delta S_{\rm KL,cycle}$, as established above. At the other extreme, the instantaneous compression protocol yields the maximal excess work $\Delta W^{\rm max}$, given by Eq.~(\ref{eq:exworkinst}), among protocols with monotonically increasing trap stiffness. Combining these results, we obtain the lower and upper bounds on the total entropy production of the resetting cycle:
\begin{equation}
\Delta S_{\rm KL,cycle}\leq\Delta S_{\rm cycle}\leq\Delta S_{\rm KL,cycle}+\frac{\Delta W^{\rm max}}{T}
\label{eq:bounds}.
\end{equation}

\subsection{Entropy production rate}
\label{sec:epr}

Having established the lower and upper bounds on the entropy
production per resetting cycle, we now turn to the entropy
production rate (EPR). The EPR is obtained by dividing the
entropy production per cycle by the total cycle duration
$t+t^*$, where $t$ is the diffusion interval and $t^*$
is the duration of the resetting process. In the quasistatic
limit, $t^*\rightarrow\infty$, the EPR vanishes despite
the finite entropy production per cycle. At the opposite
extreme, we consider instantaneous switching to the final
stiffness $k_0$, followed by extremely rapid relaxation
to the localized state. Assuming sufficiently high mobility,
the relaxation time becomes negligible compared with $t$,
so that the entire resetting process can be regarded as
instantaneous ($t^*\rightarrow0$).This limit maximizes the entropy production per cycle
while minimizing the total cycle time, and thus yields
the maximal EPR. It also provides the closest physical
realization of the teleportation picture of stochastic
resetting, with an explicitly accounted operational cost. 

During free diffusion, the variance evolves as $\sigma^2(t)=\sigma_0^2+2Dt$, where $D=k_BT/\gamma$. Using $\sigma_0^2=k_BT/k_0$ and defining the characteristic relaxation time $\tau=\gamma/k_0=\sigma_0^2/D$, we obtain the dimensionless variance ratio $R(t)\equiv\sigma^2(t)/\sigma_0^2=1+2t/\tau$. For Poissonian resetting, the diffusion intervals are exponentially distributed, $p(t)=re^{-rt}$, where $r$ is the resetting rate. Since the resetting time $t^*$ vanishes in the instantaneous limit, the mean cycle duration is $1/r$. We average separately the end-point KL contribution in Eq.~(\ref{eq:deltascycle}) and the operational contribution $\Delta S_{\rm op}=\Delta W/T$, with the operational work given by Eq.~(\ref{eq:exworkinst}), over this distribution. Dividing by the mean cycle duration, we obtain
\begin{equation}
\dot S_{\rm KL, cycle}=\frac{k_Br}{2}\left[\frac{2}{r\tau}-1+\frac{r\tau}{2}e^{r\tau/2}E_1\left(\frac{r\tau}{2}\right)\right],
\label{eq:eprkl}
\end{equation}
and
\begin{equation}
\dot S_{\rm op}=\frac{k_Br}{2}\left[\frac{2}{r\tau}-e^{r\tau/2}E_1\left(\frac{r\tau}{2}\right)\right],
\label{eq:eprop}
\end{equation}
where $E_1(x)=\int_x^\infty e^{-u}/u\,du$ is the exponential integral. The end-point KL contribution in Eq.~(\ref{eq:eprkl}) provides a lower bound on the EPR, while its sum with the operational contribution in Eq.~(\ref{eq:eprop}) gives the upper bound.

\begin{figure}[t]
    \centering
    \includegraphics[width=0.5\textwidth]{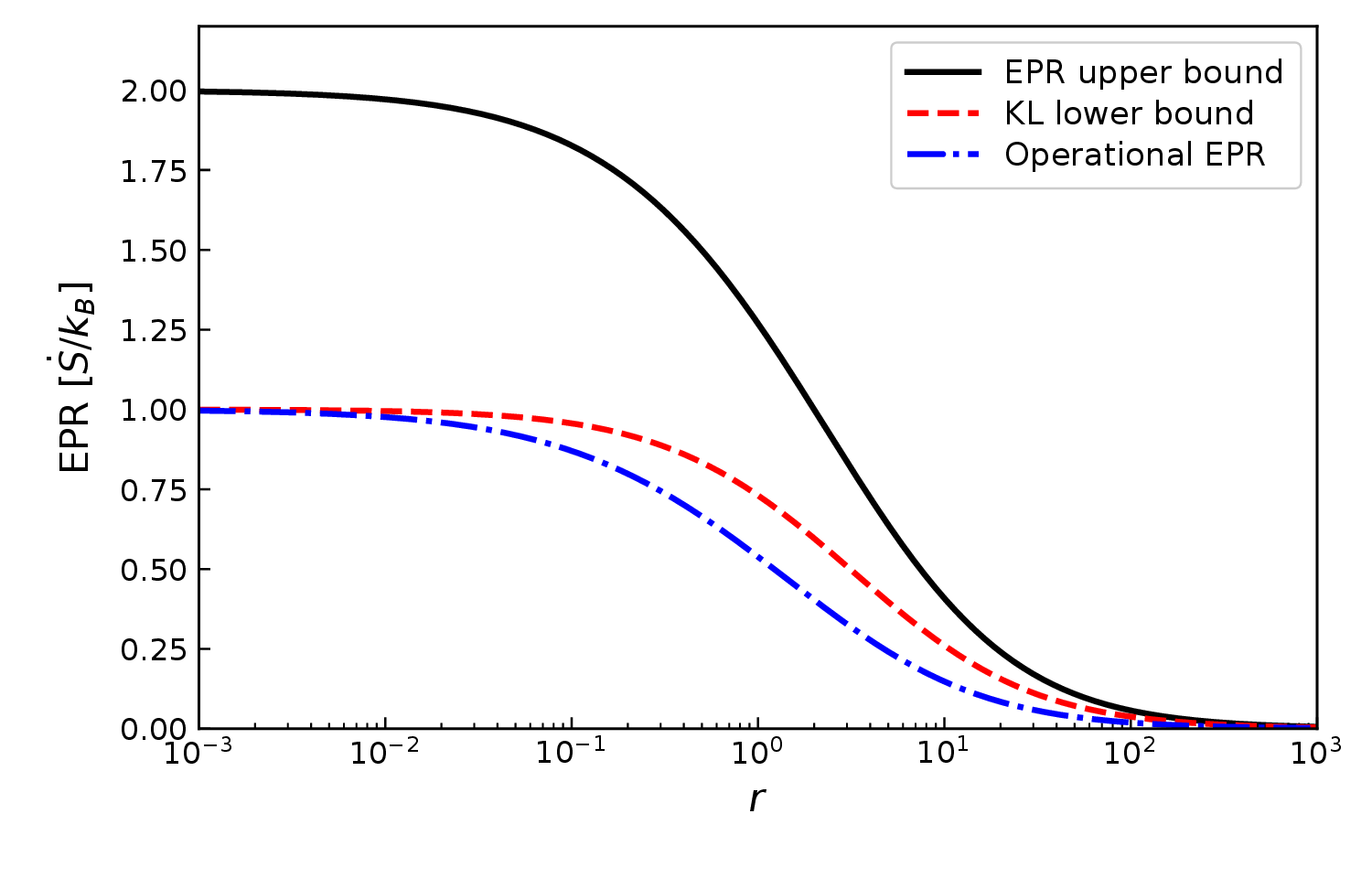}
\caption{Entropy production rate as a function of the Poissonian resetting rate $r$ for $\tau=1$. The curves assume negligible resetting duration ($t^*=0$), an approximation that breaks down at high resetting rates. The solid black curve shows the upper bound, given by the sum of the KL and operational contributions in Eqs.~(\ref{eq:eprkl}) and (\ref{eq:eprop}), respectively. The dashed red and dot-dashed blue curves show these two contributions separately. The high-rate correction is discussed in the text.} 
    \label{fig:fig1}
\end{figure}

Figure~\ref{fig:fig1} shows the EPR as a function of the resetting rate $r$ for Poissonian resetting, with $\tau=1$. The solid black curve represents the upper bound, obtained by summing the KL lower bound and the operational contribution in Eqs.~(\ref{eq:eprkl}) and (\ref{eq:eprop}), respectively. The dashed red and dot-dashed blue curves show these two contributions separately. Both contributions decrease monotonically with increasing $r$, approaching finite plateaus at low resetting rates and vanishing at high rates. This monotonic behavior is not obvious a priori, since
longer diffusion intervals produce more entropy per cycle
but also reduce the frequency of resetting events.
The competition between these effects could, in principle,
give rise to a maximum at an intermediate resetting rate.
To further characterize the observed behavior, we examine
the asymptotic limits of the EPR.

In the low-rate limit $r\tau\ll1$, we use the expansions $e^x=1+O(x)$ and $E_1(x)=-\ln x+O(1)$ for $x\to0$. These imply that the terms involving the exponential integral in Eqs.~(\ref{eq:eprkl}) and (\ref{eq:eprop}) become negligible compared to the leading term $2/(r\tau)$, yielding
\begin{equation}
\dot S_{\rm KL,cycle}\simeq\dot S_{\rm op}\simeq\frac{k_B}{\tau},
\qquad r\tau\ll1,
\label{eq:lowrate}
\end{equation}
which explains the plateaus in Fig.~\ref{fig:fig1}. The physical origin of these finite limits lies in the unbounded spreading of the particle during free diffusion. Indeed, the variance $\sigma^2(t)=\sigma_0^2+2Dt$ grows linearly with time without approaching a stationary value. As a result, both contributions to the entropy production {\em per cycle}\/ grow linearly with $t$, asymptotically compensating for the decreasing frequency of resetting events.

In the opposite high-rate limit, the condition $r\tau\gg1$ implies that $2Dt\ll\sigma_0^2$. This is the limit of {\em perpetual resetting}, in which the particle effectively remains in its initial equilibrium distribution. In this regime, the restoring drift requires approximately the same amount of time to reverse the small diffusive expansion, so that $t^*\simeq t$. Consequently, the total cycle duration approaches $2t$, rather than $t$ as assumed in deriving Eqs.~(\ref{eq:eprkl}) and (\ref{eq:eprop}). The high-rate asymptotics of these expressions must therefore be multiplied by $1/2$. Using the large-argument expansion $e^xE_1(x)=1/x-1/x^2+2/x^3+O(x^{-4})$, with $x=r\tau/2$, we obtain
\begin{equation}
\dot S_{\rm KL,cycle}\simeq\frac{2k_B}{r\tau^2},
\qquad
\dot S_{\rm op}\simeq\frac{k_B}{r\tau^2},
\qquad r\tau\gg1.
\label{eq:highrate}
\end{equation}
Thus, both contributions vanish as $r\to\infty$, with the KL contribution twice as large as the operational contribution. 


It is worth noting that the finite EPR obtained in the low-rate limit is a particular consequence of free diffusion. If the particle instead diffuses in a weak harmonic potential of stiffness $k_1$, with $0<k_1<k_0$, its variance evolves as
\begin{equation}
\sigma^2(t)=\frac{k_BT}{k_1}+\left(\sigma_0^2-\frac{k_BT}{k_1}\right)e^{-2k_1t/\gamma},
\label{eq:app_variance}
\end{equation}
approaching the finite equilibrium value $k_BT/k_1$ at long times. Consequently, both the KL and operational contributions to the entropy production per cycle saturate. Since the mean cycle duration diverges as $1/r$, both contributions to the EPR vanish as $r\to0$. As they also vanish at high resetting rates, the corresponding bounds are expected to exhibit a maximum at an intermediate rate. Thus, even arbitrarily weak harmonic confinement qualitatively changes the low-rate behavior, highlighting the singular nature of the free-diffusion limit. A detailed derivation of the entropy-production rates for a nonzero stiffness $k_1$, including their low- and high-resetting-rate asymptotic behavior, is provided in Appendix~\ref{app:finite_stiffness}

\section{Summary}

We have investigated the entropy production associated with stochastic resetting by interpreting the process as a thermodynamic cycle consisting of thermal expansion during diffusion and compression performed by an external resetting agent (an external demon). This picture allows us to distinguish between the irreversibility associated with the resetting cycle and the additional entropy production arising from the demon's intervention. The former is characterized by an information-theoretic lower bound based on the KL divergence between forward and reversed trajectory ensembles. Although this divergence generally depends on the full stochastic trajectories, for a harmonic trap it reduces to an expression involving only the probability distributions before and after resetting. The latter, which we refer to as the operational contribution, depends on the resetting protocol and is quantified here by the excess work required for instantaneous compression relative to a quasistatic protocol. Since instantaneous compression entails the largest excess work among protocols that monotonically increase the trap stiffness, it provides an upper bound on the operational entropy production. A complementary interpretation of the resetting stage can be formulated in terms of information erasure. During diffusion, the particle's positional distribution broadens, increasing its configurational entropy. The demon subsequently restores the initial distribution irrespective of the particle's position at the end of diffusion, effectively erasing information about its preceding state. This resembles Landauer erasure~\cite{Landauer1961IBM}, in which information about the preceding state is discarded.

We derived analytical expressions for the corresponding entropy-production rates and examined their dependence on the resetting rate. For free diffusion between resetting events, both contributions approach finite values in the low-rate limit and vanish at high resetting rates. The finite low-rate values arise because the increasing entropy production per cycle compensates for the decreasing frequency of resetting events, while at high resetting rates the particle has insufficient time to spread appreciably between successive resets. By contrast, even weak harmonic confinement causes the entropy-production rates to vanish as $r\to0$, since the particle's distribution eventually relaxes to equilibrium with a finite variance, causing the entropy production per cycle to saturate (see Appendix~\ref{app:finite_stiffness}).

\section*{Acknowledgements} 

I thank Saar Rahav, Yael Roichman, R\'emi Goerlich, and Ralf Metzler for their helpful comments. This work was supported by the Israel Science Foundation (ISF), Grant No. 1258/22. I also thank the support of the German Research Foundation (DFG Grant Nos. 318763901, CRC 1294 “Data Assimilation”)

\setcounter{secnumdepth}{3}
\appendix

\section{Entropy Production for Diffusion in a Harmonic Potential of Finite Stiffness}
\label{app:finite_stiffness}

We consider the case in which the particle diffuses
in a harmonic potential of stiffness $k_1$, with
$0<k_1<k_0$. Starting from the equilibrium distribution
with variance $\sigma_0^2=k_BT/k_0$, the variance
during the diffusion stage evolves according to
Eq.~(\ref{eq:app_variance}).
Introducing $\alpha=k_1/k_0$ and $\tau=\gamma/k_0$, we define
\begin{equation}
R(t)\equiv\frac{\sigma^2(t)}{\sigma_0^2}
=\frac{1-(1-\alpha)e^{-2\alpha t/\tau}}{\alpha}.
\label{eq:app_R}
\end{equation}

Since the initial and final distributions remain Gaussian, the KL lower bound, Eq.~(\ref{eq:deltascycle}), retains its form:
\begin{equation}
\Delta S_{\rm cycle}\geq\Delta S_{\rm KL,cycle}
=\frac{k_B}{2}\left[R(t)+\frac{1}{R(t)}-2\right].
\label{eq:app_KL_cycle}
\end{equation}

The operational contribution also retains the form of Eq.~(\ref{eq:exworkinst}), with $R(t)$ given by Eq.~(\ref{eq:app_R}):
\begin{equation}
\Delta S_{\rm op,cycle}
=\frac{k_B}{2}\left[R(t)-1-\ln R(t)\right].
\label{eq:app_op_cycle}
\end{equation}
This is because the initial switch from $k_1$ to $k(t)=k_BT/\sigma^2(t)$ is accounted for in the expansion branch, while the operational contribution involves only the subsequent resetting from $k(t)$ to $k_0$. The excess work between the instantaneous and quasistatic resetting protocols therefore depends only on $R(t)$.

The EPRs are obtained by the same Poissonian averaging procedure as in the main text, replacing the free-diffusion variance ratio by Eq.~(\ref{eq:app_R}):
\begin{align}
\dot S_{\rm KL}
&=\frac{k_Br^2}{2}\int_0^\infty e^{-rt}
\left[R(t)+\frac{1}{R(t)}-2\right]dt,
\label{eq:app_KL_rate}\\
\dot S_{\rm op}
&=\frac{k_Br^2}{2}\int_0^\infty e^{-rt}
\left[R(t)-1-\ln R(t)\right]dt.
\label{eq:app_op_rate}
\end{align}
In the high-rate limit, $r\tau\gg1$, the short-time expansion of Eq.~(\ref{eq:app_R}) gives
\begin{equation}
R(t)=1+\frac{2(1-\alpha)t}{\tau}+O(t^2).
\end{equation}
Substituting into Eqs.~(\ref{eq:app_KL_rate}) and (\ref{eq:app_op_rate}), and including the factor of $1/2$ (see discussion of the high rate limit is section~\ref{sec:epr}), we obtain
\begin{equation}
\dot S_{\rm KL}\simeq\frac{2k_B(1-\alpha)^2}{r\tau^2},
\qquad
\dot S_{\rm op}\simeq\frac{k_B(1-\alpha)^2}{r\tau^2},
\end{equation}
which recovers the high-rate asymptotic results of Eq.~(\ref{eq:highrate})  as $\alpha\to0$.

\begin{figure}[t]
    \centering
    \includegraphics[width=0.5\textwidth]{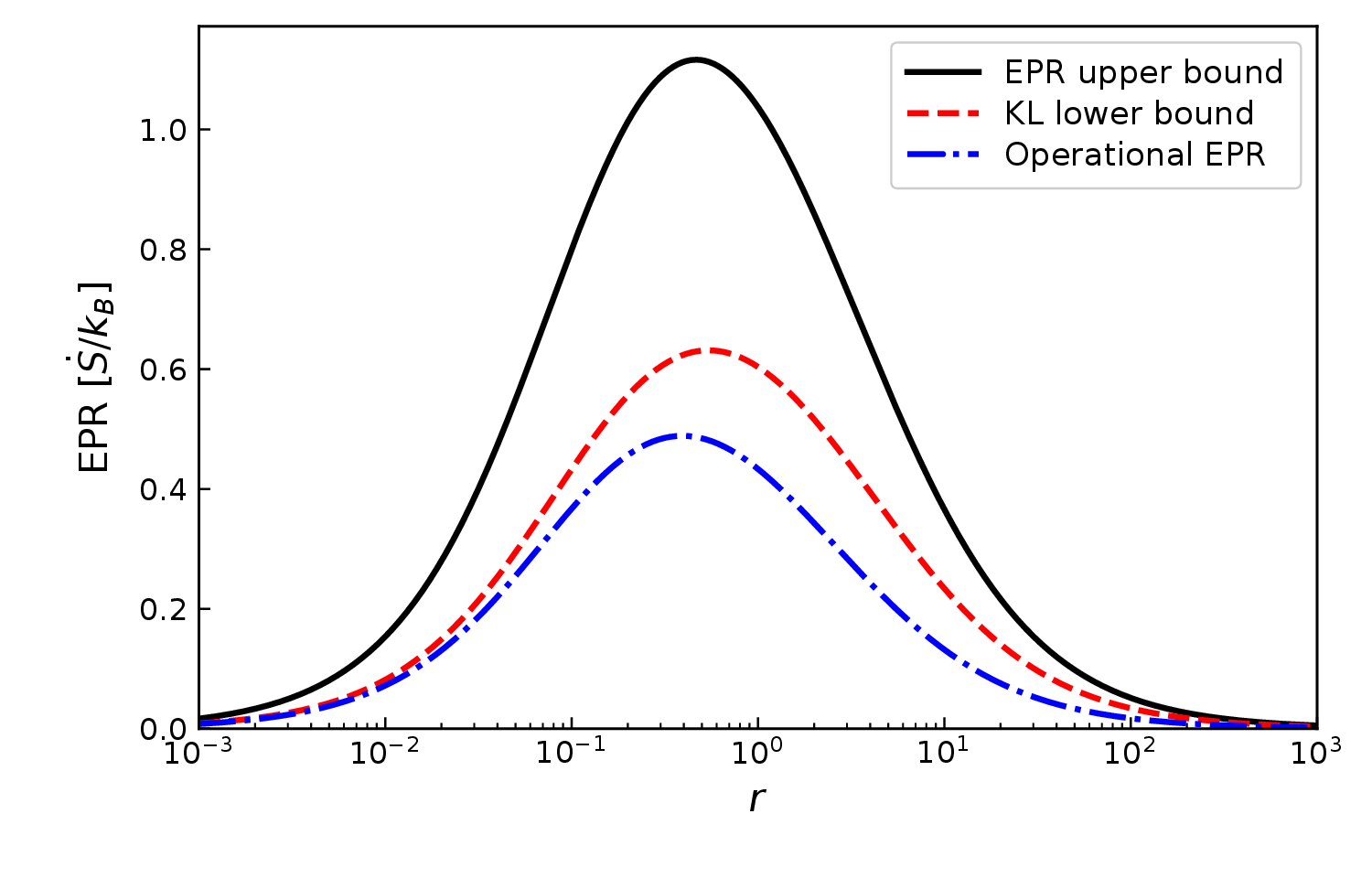}
\caption{Entropy-production rates as functions of the resetting rate $r$ for diffusion in a harmonic potential with stiffness $k_1=0.05k_0$ ($\tau=1$). The solid black curve shows the EPR upper bound, the dashed red curve the KL lower bound, and the dash-dotted blue curve the operational contribution.}
\label{fig:finite_stiffness}
    \label{fig:fig2}
\end{figure}

The low-rate behavior is more subtle because finite stiffness $k_1$ introduces an additional relaxation timescale, $\gamma/k_1=\tau/\alpha$. While the condition $r\tau\ll1$ defines the low-rate regime for free diffusion, saturation of the variance before resetting requires $r\tau\ll2\alpha$, which is more restrictive for weak confinement ($\alpha\ll1$). In this regime, $R(t)$ approaches its stationary value $1/\alpha$ over a typical diffusion interval, so that the entropy production per cycle remains finite even as $r\to0$. In this limit, the EPRs become
\begin{equation}
\dot S_{\rm KL}\simeq\frac{k_Br}{2}
\left(\frac{1}{\alpha}+\alpha-2\right),
\quad
\dot S_{\rm op}\frac{k_Br}{2}
\left(\frac{1}{\alpha}-1+\ln\alpha\right).
\end{equation}
Both contributions vanish linearly with $r$, in contrast to the finite low-rate limits obtained for free diffusion, reflecting the noncommutativity of the limits $r\to0$ and $\alpha\to0$. Since the EPRs also vanish at high resetting rates, both exhibit maxima at intermediate values of $r$. This behavior is illustrated in Fig.~\ref{fig:fig2} for $\alpha=0.05$, where the EPR upper bound reaches a maximum of approximately $\dot S/k_B=1.12$ at $r=0.47$ (with $\tau=1$).

\bibliography{main}

\end{document}